\documentclass[a4paper, 11pt]{article}
\pdfoutput=1

\usepackage[british]{babel}
\usepackage[T1]{fontenc}
\usepackage[utf8]{inputenc}
\usepackage{jheppub}
\usepackage{microtype}
\usepackage{booktabs}
\usepackage{capt-of}
\usepackage{xfrac}
\usepackage{mleftright}
\usepackage{siunitx}
\usepackage{csquotes}
\usepackage[noabbrev]{cleveref}

\DeclareMathOperator{\Tr}{Tr}
\DeclareMathOperator{\diag}{diag}
\DeclareSIUnit\pc{pc}

\newcommand*{\tran}{{\mathsf{T}}}
\newcommand*{\lr}{\mleft(}
\newcommand*{\rr}{\mright)}

\newcommand*{\SUIIIC}{\operatorname{SU}(3)_{\text{C}}}
\newcommand*{\SUIIL}{\operatorname{SU}(2)_{\text{L}}}
\newcommand*{\UIY}{\operatorname{U}(1)_{\text{Y}}}

\newcommand*{\OmegacObs}{\Omega_{\text{c}}^{\text{obs}}}

\hypersetup{
	pdftitle={Singlet--doublet fermion and triplet scalar dark matter with radiative neutrino masses},
	pdfauthor={Juri Fiaschi, Michael Klasen, Simon May},
	pdfkeywords={Beyond Standard Model, Cosmology of Theories beyond the SM, Neutrino Physics, Discrete Symmetries}
}

\preprint{MS-TP-18-33}

\arxivnumber{1812.11133}

\title{Singlet--doublet fermion and triplet scalar dark matter with radiative neutrino masses}

\author[a]{Juri Fiaschi,}
\author[a]{Michael Klasen}
\author[a, b]{and Simon May}

\affiliation[a]{Institut für Theoretische Physik, Westfälische Wilhelms-Universität
 Münster, Wilhelm-Klemm-Straße 9, 48149 Münster, Germany}
\affiliation[b]{Max-Planck-Institut für Astrophysik, Karl-Schwarzschild-Straße 1, 85741 Garching, Germany}

\emailAdd{fiaschi@uni-muenster.de}
\emailAdd{michael.klasen@uni-muenster.de}
\emailAdd{simon.may@uni-muenster.de}

\abstract{%
  We present a detailed study of a combined singlet--doublet fermion and
  triplet scalar model for dark matter. These models have only
  been studied separately in the past. Together, they form a simple
  extension of the Standard Model that can account for dark matter and
  explain the existence of neutrino masses, which are generated radiatively.
  This holds even if singlet--doublet fermions and triplet scalars never contribute simultaneously to the dark matter abundance.
  However, this also implies the existence of lepton flavour violating
  processes. In addition, this particular model allows for gauge coupling
  unification. The new fields are odd under a new $\mathbb{Z}_2$ symmetry
  to stabilise the dark matter candidate. We analyse the dark matter,
  neutrino mass and lepton flavour violation aspects both separately and in conjunction,
  exploring the viable parameter space of the model. This is done using
  a numerical random scan imposing successively the neutrino
  mass and mixing, relic density, Higgs mass, direct detection, collider and lepton
  flavour violation constraints. We find that dark matter in this model is
  fermionic for masses below about \SI{1}{\TeV} and scalar above. The narrow mass
  regions found previously for the two separate models are enlarged by
  their coupling. While coannihilations of the weak isospin partners are
  sizeable, this is not the case for fermions and scalars despite their
  often similar masses due to the relatively small coupling of the two sectors,
  imposed by the small neutrino masses. We observe a high degree of
  complementarity between direct detection and lepton flavour violation
  experiments, which should soon allow to fully probe the fermionic dark
  matter sector and at least partially the scalar dark matter sector.%
}


\begin{document}
\maketitle
\flushbottom

\section{Introduction}

Evidence for dark matter (DM) is accumulating from astrophysical observations
at many different length scales \cite{Klasen:2015uma}. Since no particle in
the Standard Model (SM) of particle physics has the required properties of
being uncharged under the electromagnetic and strong interactions and to be
sufficiently cold to explain the observed large scale structure in the Universe,
DM is widely believed to provide a strong indication for physics beyond the SM.

The simplest solution to the DM puzzle consists in adding a small number of new particles
to the SM, out of which the lightest is neutral and stable, e.\,g.\ due to a
discrete $\mathbb{Z}_2$ symmetry. Weakly interacting massive particles (WIMPs), such as
heavy scalars or neutrinos, are among the most promising candidates, as they
have the right cross section to be produced as thermal relics with the observed
density $\OmegacObs h^2 = \num{0.120 +- 0.001}$ \cite{Aghanim:2018eyx}. Here, $h$ denotes the
present Hubble expansion rate in units of \SI{100}{\km\per\s\per\mega\pc}.
Many of these so-called minimal models have been studied in the past, including
those with scalar singlets
\cite{Burgess:2000yq}, doublets \cite{LopezHonorez:2006gr,Klasen:2013btp},
and triplets \cite{Araki:2011hm,Fischer:2011zz,JosseMichaux:2012wj,%
Ayazi:2015mva,Khan:2016sxm},
fermion singlets \cite{Esch:2013rta,Klasen:2013ypa}, doublets
\cite{Cohen:2011ec,Cheung:2013dua,Calibbi:2015nha}, and triplets \cite{Dedes:2014hga},
as well as with higher scalar \cite{Hambye:2009pw}
and/or fermion multiplets \cite{Cirelli:2005uq}. In
particular, triplet scalar DM requires hypercharge $Y = 0$ such that the $\SUIIL$
component with $T_3 = 0$ is neutral \cite{Restrepo:2013aga}.
On the other hand, mixing fermion singlets and
doublets reduces the coupling to weak gauge bosons and can transform DM from a
Dirac into a Majorana particle, yielding both the correct relic density and still
allowed direct detection cross sections \cite{Cohen:2011ec}.

Particularly well motivated models are those that do not only provide a DM candidate,
but also solve other SM problems such as the smallness of neutrino masses.
This is possible when the $d = 5$ Weinberg operator is realised at one loop
\cite{Bonnet:2012kz}, such that the particles in the loop have opposite $\mathbb{Z}_2$
parity to the SM particles and include a neutral DM candidate \cite{Restrepo:2013aga}.
One of the most popular so-called radiative seesaw models is the scotogenic
model with only one additional scalar (inert Higgs) $\SUIIL$ doublet
and a (right-handed neutrino) fermion singlet \cite{Ma:2006km}, for
which we recently demonstrated the importance of coannihilations
between the scalar dark matter and the right-handed neutrinos
\cite{Klasen:2013jpa}. Many variants with general multiplets
containing two scalars and one fermion have subsequently
been proposed. In particular, the observed relic density was found to require
a scalar triplet DM mass of \SI{2.5}{\TeV} and couplings of $\mathcal{O}(\num{e-3})$ to an
additional charged scalar singlet and vector-like doublet fermion to reproduce
the observed neutrino mass differences 
\cite{Law:2013saa,Brdar:2013iea}, whereas the scalar triplet DM mass could vary
from \SIrange{1.7}{2.6}{\TeV} when a neutral scalar singlet and vector-like doublet
fermion were added \cite{Farzan:2010mr,Law:2013saa}. 
Unfortunately, these masses/couplings are too large/small to explain
the observed tension in the muon's $g-2$ between theory and experiment
\cite{Okada:2015vwh}.

The models above all belong to the one-loop topology T3 with two scalars
and one fermion \cite{Bonnet:2012kz,Restrepo:2013aga}. In this paper, we
study a model of topology T1-3 with one scalar and two fermions, one of
which is vector-like. In contrast to the first of these models (T1-3-A
with hypercharge parameter $\alpha = 0$), where the scalar DM was a
singlet, had to be lighter than \SI{600}{\GeV}, and where the parameter space
will soon be explored by $\mu \to 3e$ and $\mu$--$e$ conversion experiments
\cite{Klasen:2016vgl}, we investigate here a model (T1-3-B, also with
$\alpha = 0$) where the scalar DM is the neutral component of a triplet.
In both cases, (at least) two generations of scalars must be introduced,
so that not only the constraints from the DM relic density and observed Higgs
mass, but also two non-zero neutrino mass differences
and mixing angles can be imposed \cite{Tanabashi:2018oca}. As for the
model T1-3-A,
we restrict ourselves to two scalar generations, which implies that the
lightest neutrino is massless. Both models,
like a previously studied model with both singlet--doublet scalars and
fermions (T1-2-A with $\alpha = 0$) \cite{Esch:2018ccs}, have the additional
advantage that they allow for gauge coupling unification at a scale
$\Lambda = \mathcal{O}(\SI{e13}{\GeV})$, at variance with most of the other
models mentioned above including the original scotogenic model
\cite{Hagedorn:2016dze}.
Like all models classified in \cite{Restrepo:2013aga}, these models feature Majorana neutrinos.

From the list of models with DM, radiative neutrino masses and gauge coupling
unification which have not been studied previously, our model features the
smallest number (six) of new interaction terms, which makes it possible to fully
explore its parameter space. We do so by constructing the most general Lagrangian
from the field content, eliminating a redundant term, generating model files
for \texttt{SPheno} and \texttt{micrOMEGAs}
with \texttt{SARAH}~4.13.0 \cite{Staub:2013tta}, imposing the neutrino masses using the
Casas--Ibarra parametrisation \cite{Casas:2001sr}, and calculating the physical
particle spectrum and relevant precision observables with \texttt{SPheno}~4.0.3
\cite{Porod:2011nf} as well as the DM relic density and direct detection cross
sections with \texttt{micrOMEGAs}~4.3.5 \cite{Barducci:2016pcb}.
The constructed Lagrangian was checked and template \texttt{SARAH} model files
were generated by the newly-developed tool \texttt{minimal-lagrangians}, which is
available upon request.

The remainder of this paper is organised as follows: In \cref{sec:model} we
introduce the model under examination, giving details of its field content and
the full Lagrangian, describing the mixing of fundamental particles into
physical states in the fermion and scalar sectors as well as the mass
splitting between neutral and charged states. In \cref{sec:dm_relic}, we discuss
the features of the two distinct DM candidates within our model and update
and clarify contradictory results in the literature. In \cref{sec:radiative_m_nu} we
connect the dark sector to the radiative generation of neutrino masses and
explain how we impose neutrino constraints on the parameter space of the model.
In \cref{sec:dd_lfv} we study their impact on the nature, parameter space and
direct detection prospects of the DM candidates and make predictions for
the automatically generated lepton flavour violating (LFV) processes.
Our conclusions are given in \cref{sec:conclusion}.


\section{Description of the model}
\label{sec:model}

\begin{figure}
	\centering
	\includegraphics{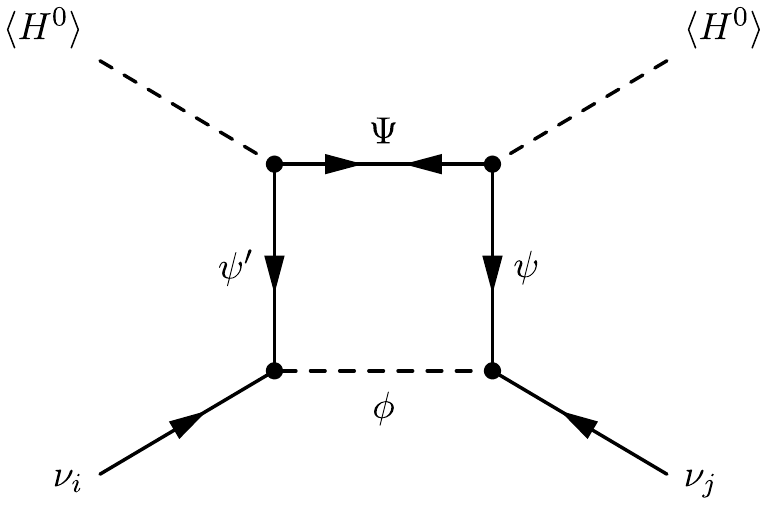}
	\caption{%
		Loop topology T1-3 for the radiative generation of neutrino masses at the one-loop level.%
	}
	\label{fig:feynman_diagram_m_nu_before_mixing}
\end{figure}

\begin{table}
	\centering
	\caption{New fields and their quantum numbers in the model T1-3-B with $\alpha = 0$.}
	\label{tab:model_content}
	\medskip
	\begin{tabular}{c c c c c c c c}
		\toprule
		Field & Generations & Spin & Lorentz rep. & $\SUIIIC$ & $\SUIIL$ & $\UIY$ & $\mathbb{Z}_2$\\
		\midrule
		$\Psi$ & 1 & $\sfrac{1}{2}$ & $(\sfrac{1}{2}, 0)$ & $\mathbf{1}$ & $\mathbf{1}$ & $0$ & $-1$ \\
		$\psi$ & 1 & $\sfrac{1}{2}$ & $(\sfrac{1}{2}, 0)$ & $\mathbf{1}$ & $\mathbf{2}$ & $-1$ & $-1$ \\
		$\psi'$ & 1 & $\sfrac{1}{2}$ & $(\sfrac{1}{2}, 0)$ & $\mathbf{1}$ & $\mathbf{2}$ & $1$ & $-1$ \\
		$\phi_i$ & 2 & $0$ & $(0, 0)$ & $\mathbf{1}$ & $\mathbf{3}$ & $0$ & $-1$ \\
		\bottomrule
	\end{tabular}
\end{table}

Following the notation in ref.\ \cite{Bonnet:2012kz,Restrepo:2013aga},
the model T1-3-B with hypercharge parameter $\alpha = 0$ is defined by
extending the SM with the colour-singlet fields in \cref{tab:model_content}.
T1-3 labels the general one-loop topology generating radiative neutrino masses,\footnote{%
	The three topologies T1-1, T1-2 and T1-3, collectively T1-$n$, differ only in the number $n$ of fermion lines (as opposed to scalar lines) in the loop.%
}
as shown in \cref{fig:feynman_diagram_m_nu_before_mixing}, while T1-3-B is a specific $\SUIIL$ multiplet structure of the new fields.
The parameter $\alpha$ is simply a common offset to the hypercharge values given in \cref{tab:model_content} -- for example, in the model T1-3-B with $\alpha = 2$, every value in the column $\UIY$ is increased by \num{2} compared to T1-3-B with $\alpha = 0$.

The present model therefore combines the $\SUIIL$ triplet scalars $\phi_i$ of zero hypercharge
($\UIY$) with an $\SUIIL$ singlet fermion $\Psi$ and doublet fermions $\psi, \psi'$. They are
all odd under the discrete symmetry $\mathbb{Z}_2$, while the SM fields are even.
The components of the new fields are given by
\begin{equation}
	\Psi = \Psi^0
	,\quad
	\psi = \begin{pmatrix} \psi^0 \\ \psi^- \end{pmatrix}
	,\quad
	\psi' = \begin{pmatrix} \psi'^+ \\ \psi'^0 \end{pmatrix}
	,\quad
	\phi_i = \begin{pmatrix}
		\frac{1}{\sqrt{2}} \phi_i^0 & \phi_i^+ \\
		\phi_i^- & -\frac{1}{\sqrt{2}} \phi_i^0
	\end{pmatrix},
\end{equation}
where superscripts indicate electric charges. To obtain two non-zero neutrino
mass differences, two generations of scalar triplets are required. Since
the scalar triplets have zero hypercharge, they are treated as real,
$(\phi^0_i)^\dagger = \phi^0_i, (\phi^+_i)^\dagger = \phi^-_i$. $\Psi$ has the same
quantum numbers as a $\mathbb{Z}_2$-odd right-handed neutrino, whereas $\psi$
and $\psi'$ together form a $\mathbb{Z}_2$-odd vector-like lepton doublet, which
makes the model automatically anomaly-free. In principle, all neutral field
components are possible DM candidates. The $\mathbb{Z}_2$ symmetry not only
stabilises the lightest new particle against decay into SM fields, but also
forbids neutrino masses from a tree-level type-I seesaw mechanism.

The most general renormalisable Lagrangian for the model is\footnote{%
	Here and in the following, products of field multiplets are always defined to be
	Lorentz- and gauge-invariant. For example, the product of the two $\SUIIL$
	Weyl spinors $\psi$ and $\psi'$ should be interpreted as
	$\psi \psi' = \varepsilon^{ji} \varepsilon^{j'i'} \psi_{ii'} \psi'_{jj'}$, where the indices represent the $\SUIIL$ and spinor components.%
}
\begin{align}
 \mathcal{L} &= \mathcal{L}_{\text{SM}} + \mathcal{L}_{\text{kin}}
 - \frac{1}{2} (M_\phi^2)^{ij} \Tr(\phi_i \phi_j)
 - \lr \frac{1}{2} M_\Psi \Psi \Psi + \text{H.\,c.} \rr
 - \lr M_{\psi\psi'} \psi \psi' + \text{H.\,c.} \rr \nonumber\\
 &\quad
 - (\lambda_2)^{ij} H^\dagger \phi_i \phi_j H
 - (\lambda_3)^{ijkm} \Tr(\phi_i \phi_j \phi_k \phi_m) \nonumber\\
 \label{eq:lagrangian}
 &\quad
 - \lr \lambda_4 (H^\dagger \psi') \Psi + \text{H.\,c.} \rr
 - \lr \lambda_5 (H \psi) \Psi + \text{H.\,c.} \rr
 - \lr (\lambda_6)^{ij} L_i \phi_j \psi' + \text{H.\,c.} \rr,
\end{align}
where $H$ is the SM Higgs field (with vacuum expectation value $v$ and quartic coupling $\lambda$)\footnote{%
	We have used a value of $\lambda = \num{0.2612}$ as an input to \texttt{SPheno} throughout the analysis.%
}
and we have combined the term
\begin{equation}
	\label{eq:lagrangian_lambda1}
	-(\lambda_1)^{ij} (H^\dagger H) \Tr(\phi_i \phi_j)
\end{equation}
with the one proportional to $\lambda_2$ using the identity
\begin{equation}
	H^\dagger \{\phi_i, \phi_j\} H = (H^\dagger H) \Tr(\phi_i \phi_j).
\end{equation}
However, for the remaining analysis, we will instead use the $\lambda_1$ term \eqref{eq:lagrangian_lambda1}
for easier comparison with the previous literature on models with one scalar triplet,
which is completely equivalent as long as there are no CP-violating phases.
The couplings $\lambda_4$ and $\lambda_5$ have the function of Yukawa terms,
which link the fermion singlet and doublets to the SM Higgs boson.
The coupling $\lambda_6$ connects the SM lepton doublet $L$ to the new fields, so
that it will be involved in the process of radiative neutrino masses generation.\footnote{%
	Except for $(M_\phi^2)^{ij}, \lambda_1^{ij}, \lambda_3^{ij} \in \mathbb{R}$,
	the parameters can, in principle, have a complex CP-violating phase.
	Note that not all of the parameters are independent: $(M_\phi^2)^{ij} = (M_\phi^2)^{ji}$,
	$(\lambda_1)^{ij} = (\lambda_1)^{ji}$, ${(\lambda_2)^{ij}}^* = (\lambda_2)^{ji}$,
	$(\lambda_3)^{ijkm} = (\lambda_3)^{jkmi} = (\lambda_3)^{kmij} = (\lambda_3)^{mijk}$.%
}

In order to ensure that the theory has a stable vacuum, the scalar potential in \eqref{eq:lagrangian} should have a minimum, i.\,e.\ it should be bounded from below.
For the SM Higgs field with a single (inert) scalar triplet (one generation), constraints on the parameters following from this requirement are given in \cite{Hambye:2009pw,Araki:2011hm}.
In general, determining whether a function is bounded from below is not a trivial task and expressing the constraints on the parameters as analytic conditions is usually only possible for specific directions in the variables.
Constraints analogous to those for one generation can be obtained:
\begin{equation}
	\lambda_3^{iiii} \ge 0
	,\quad
	2 \sqrt{\lambda \lambda_3^{iiii}} \ge -\lambda_1^{ii}.
\end{equation}
In the direction where all fields are taken to infinity equally with only some possible signs between them, the condition
\begin{equation}
	\lambda + \lambda_1^{11} + \lambda_1^{22} + \lambda_3^{1111} + \lambda_3^{2222} + \lambda_3^{1212} + 2 \lambda_3^{1112} \ge 2 |\lambda_1^{12}| + 4 (|\lambda_3^{1112}| + |\lambda_3^{1222}|)
\end{equation}
must hold.
For $H = 0$, the potential has the same form as that in \cite{Gunion:2002zf} (with $d = 0$), so additional conditions can be obtained analogously, such as:
\begin{equation}
	\begin{gathered}
		\sqrt{\lambda_3^{1111} \lambda_3^{2222}}
		\ge \lambda_3^{1212} - 2\lambda_3^{1122}
		,
		\\
		\lambda_3^{1111} + \lambda_3^{2222} + \lambda_3^{1212} + 2 \lambda_3^{1122}
		\ge 4 (|\lambda_3^{1112}| + |\lambda_3^{1222}|)
		,
		\\
		\lambda_3^{1111} \lambda_3^{2222} + \frac{1}{2} \sqrt{\lambda_3^{1111} \lambda_3^{2222}} \lr \lambda_3^{1212} + 2 \lambda_3^{1122}\rr
		\ge 2 (\lambda_3^{1111} |\lambda_3^{1112}| + \lambda_3^{2222} |\lambda_3^{1222}|)
		.
	\end{gathered}
\end{equation}
To summarize, vacuum stability can be controlled through the parameter $\lambda_3$, which does however not affect the DM phenomenology (see below).
Apart from loop corrections, the vacuum structure is otherwise unchanged since the imposed exact $\mathbb{Z}_2$ symmetry prevents the new scalars $\phi_i$ from acquiring nonzero vacuum expectation values.

After electroweak symmetry breaking, the mass terms of the neutral fermions take the form
\begin{equation}
	\mathcal{L}_{\text{f},0} = -\frac{1}{2} M_\Psi \Psi \Psi - M_{\psi\psi'} \psi^0 \psi'^0
	-\frac{\lambda_4 v}{\sqrt{2}} \psi'^0 \Psi - \frac{\lambda_5 v}{\sqrt{2}} \psi^0 \Psi
	+ \text{H.\,c.},
\end{equation}
where $v$ is the vacuum expectation value of the SM Higgs field $H$.
Diagonalisation of the mass matrix \cite{Cohen:2011ec,Cheung:2013dua}
\begin{equation}
 M_{\text{f},0} = \begin{pmatrix} M_\Psi & \frac{\lambda_5 v}{\sqrt{2}} & \frac{\lambda_4 v}{\sqrt{2}} \\ \frac{\lambda_5 v}{\sqrt{2}} & 0 & M_{\psi\psi'} \\ \frac{\lambda_4 v}{\sqrt{2}} & M_{\psi\psi'} & 0\end{pmatrix}
\end{equation}
with the unitary mixing matrix $U_\chi$ allows then to rewrite the neutral mass Lagrangian
\begin{equation}
 \mathcal{L}_{\text{f},0} = -\frac{1}{2} (\Psi, \psi^0, \psi'^0) M_{\text{f},0}
 \begin{pmatrix}\Psi\\ \psi^0 \\ \psi'^0\end{pmatrix} + \text{H.\,c.}
 = -\frac{1}{2} (\chi^0)^\tran U_\chi^* M_0 U_\chi^\dagger \chi^0 + \text{H.\,c.}
\end{equation}
with
\begin{equation}
 U_\chi^* M_{\text{f},0} U_\chi^\dagger = \diag(m_{\chi_1^0}, m_{\chi_2^0}, m_{\chi_3^0})
\end{equation}
in terms of the physical Majorana fermions
\begin{equation}
 \chi^0 = U_\chi \begin{pmatrix}\Psi^0 \\ \psi^0 \\ \psi'^0\end{pmatrix}.
\end{equation}

For the two generations of scalar triplets, electroweak symmetry breaking induces
a shift in the mass matrices
\begin{equation}
 M_{\phi^0}^2 = M_{\phi^\pm}^2 = M_\phi^2 + \lambda_1 v^2.
\end{equation}
Note that at tree level, the charged and neutral mass matrices are identical.
Similarly to the neutral fermions, the scalar mass eigenstates
\begin{equation}
 \eta^{0,\pm} = O_\eta \begin{pmatrix} \phi_1^{0,\pm} \\ \phi_2^{0,\pm} \end{pmatrix}
\end{equation}
are obtained by
diagonalising the scalar mass matrices with (in this case orthogonal) rotation
matrices $O_\eta$, leading to
\begin{equation}
 O_\eta M_{\phi^{0,\pm}}^2 O_\eta^\tran = \diag(m_{\eta_1^{0,\pm}}^2, m_{\eta_2^{0,\pm}}^2).
\end{equation}
At the one-loop level, gauge interactions introduce mass splittings
\begin{equation}
 \Delta m_{\eta_i} = m_{\eta_i^{\pm}} - m_{\eta_i^0} = \SI{166}{\MeV},
\end{equation}
which, together with the new interactions,
render the neutral components slightly lighter
than the corresponding charged ones \cite{Cirelli:2005uq},
making them potential dark matter candidates.


\section{Dark matter relic density}
\label{sec:dm_relic}

Depending on the spin of the lightest neutral $\mathbb{Z}_2$-odd mass
eigenstate, our model allows for both singlet--doublet fermion and triplet
scalar DM. As mentioned in the introduction, these models have been
studied separately in the past, which, however, prohibits the radiative
generation of neutrino masses. In this section, we briefly comment on
both scenarios individually, updating and clarifying the results in the
literature.
We will limit ourselves to one generation of scalar triplets in this section,
which can be viewed as decoupling the second generation via
\begin{equation}
	(M_\phi^2)^{22} = (\SI{1000}{\TeV})^2
	,\qquad
	(M_\phi^2)^{12} =
	(\lambda_1)^{i2} =
	(\lambda_6)^{i2} = 0,
\end{equation}
since the scalars are decoupled in \cref{sec:dm_relic_fermion} anyway and the additional generation
does not introduce any qualitatively new phenomena in \cref{sec:dm_relic_scalar}.

\subsection{Singlet--doublet fermions}
\label{sec:dm_relic_fermion}

\begin{figure}
	\centering
	\includegraphics[width=\textwidth]{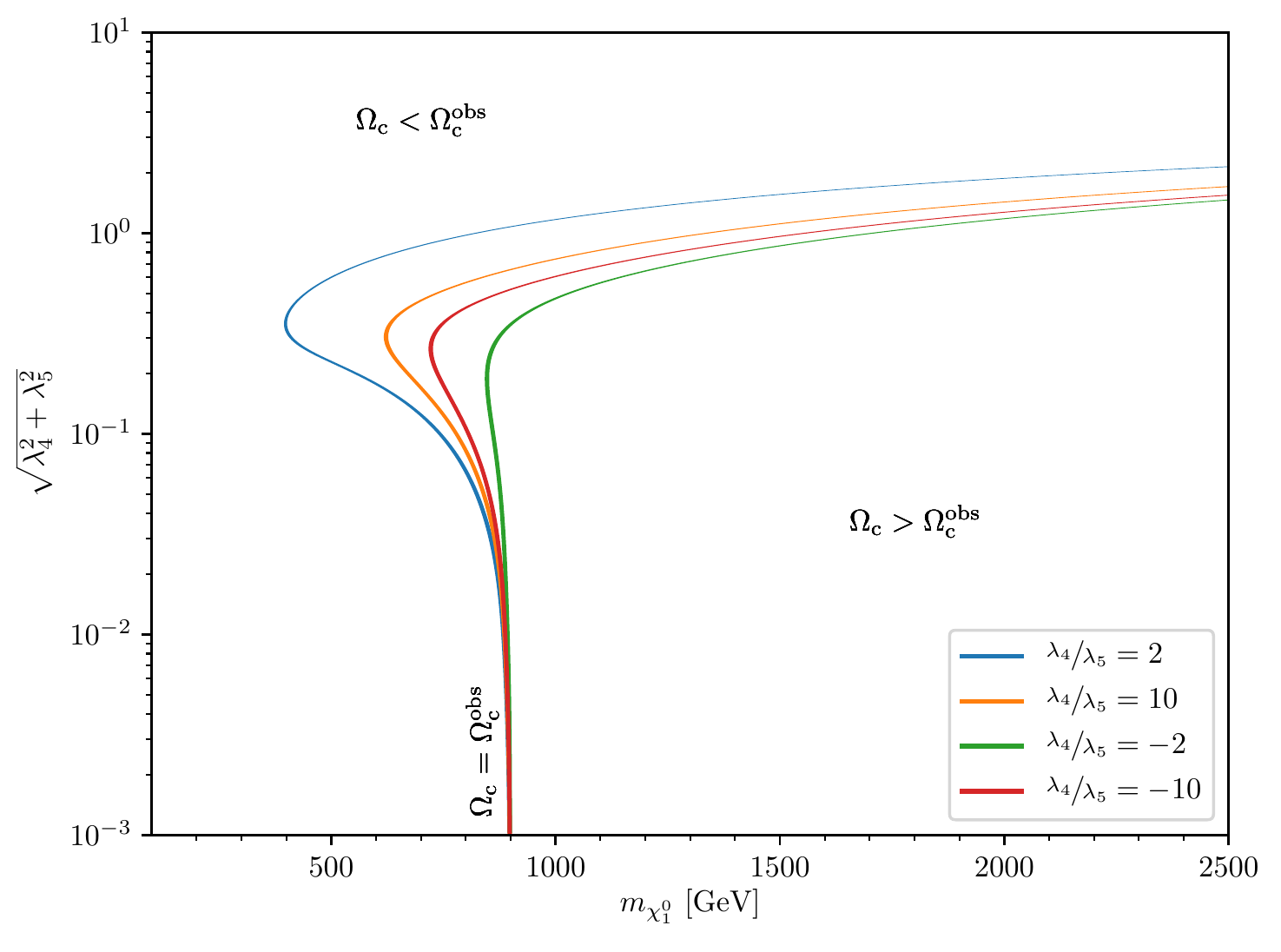}
	\caption{%
		The observed relic density contours ($\OmegacObs h^2$), separating regions of too small and too large relic density, as a function of the singlet--doublet fermion DM mass $m_{\chi_1^0}$ and the Yukawa couplings $\sqrt{\lambda_4^2 + \lambda_5^2}$ assuming $M_\Psi = M_{\psi\psi'}$ for different values of $\sfrac{\lambda_4}{\lambda_5}$.%
	}
	\label{fig:singlet_doublet_fermion_relic}
\end{figure}

We first assume that DM is composed of the lightest neutral fermion
($\chi_1^0$). The scalar sector is decoupled by setting
\begin{equation}
	(M_\phi^2)^{11} = (\SI{1000}{\TeV})^2
	,\qquad
	\lambda_1 = \lambda_3 = \lambda_6 = 0.
\end{equation}
Choosing $M_\Psi = \SI{200}{\GeV}$, $M_{\psi\psi'} = \SI{300}{\GeV}$ and $\lambda_5 = \num{0.36}$ and
neglecting loop corrections, we generally reproduce
the relic density and direct detection cross sections in figure~2 of
ref.~\cite{Cohen:2011ec}. In particular, depending on the value of $\lambda_4$ and
thus the singlet--doublet fermion mixing, blind spots of spin-independent and
-dependent direct detection experiments with no DM couplings to Higgs and
$Z$ bosons appear. Note, however, that our relic density exhibits a slightly
higher peak value when the (now known) Higgs mass of \SI{125.08}{\GeV}
\cite{Tanabashi:2018oca} is used. Also, our spin-independent direct detection
cross section is smaller by about a factor of two, which we can trace to updated
nuclear form factors in \texttt{micrOMEGAs} (version~4.3.5 vs.~2.4). A similar observation
also holds for the fermion mass scans in figure~5 of ref.~\cite{Cheung:2013dua},
where $\sfrac{\lambda_4}{\lambda_5} \in \{\pm 2, \pm 10\}$ and
$\sqrt{\lambda_4^2 + \lambda_5^2} = \num{0.3}$.
While we reproduce the relic density and spin-dependent detection contours, our
spin-independent detection contours are shifted by about \SI{0.1}{\TeV} even after
adjusting for the nuclear form factors. Furthermore, the latest XENON1T results
\cite{Aprile:2018dbl} almost double the size of the excluded low-mass regions,
so that, e.\,g., $M_\Psi \simeq M_{\psi\psi'} > \SI{1}{\TeV}$ for $\sfrac{\lambda_4}{\lambda_5} = 2$.
Below masses of \SI{1}{\TeV}, the observed relic density requires $M_\Psi\simeq M_{\psi\psi'}$
independently of $\sfrac{\lambda_4}{\lambda_5}$. For higher singlet masses, DM becomes a pure
doublet with constant mass \SI{1}{\TeV}. For other, in particular negative values of
$\sfrac{\lambda_4}{\lambda_5}$, even the new XENON1T limits become as weak as \SI{200}{\GeV}.

\Cref{fig:singlet_doublet_fermion_relic} shows another example illustrating the behavior of the singlet--doublet fermion model for the special case $M_\Psi = M_{\psi\psi'}$.
For small Yukawa couplings, both mixing between the neutral fermions and cross sections for annihilation into gauge bosons (through the mixing angles) and the Higgs boson are small.
On the other hand, for $\sqrt{\lambda_4^2 + \lambda_5^2} \ge \num{0.4}$, the DM mass must rise quickly to compensate the increasingly efficient annihilation channels.
The mass splitting depends on the relative size of the couplings $\sfrac{\lambda_4}{\lambda_5}$, controlling the impact of coannihilation processes.

\subsection{Triplet scalars}
\label{sec:dm_relic_scalar}

The literature in the case when DM is composed of the lightest neutral scalar
($\eta_1^0$) is unfortunately contradictory, and we take the opportunity to
clarify the situation. We therefore now decouple the fermion sector by setting
\begin{equation}
	M_\Psi = M_{\psi\psi'} = \SI{1000}{\TeV}
	,\qquad
	\lambda_4 = \lambda_5 = \lambda_6 = 0.
\end{equation}
Furthermore, we find that the parameter
$\lambda_3$ inducing scalar co-annihilations and conversions does not play an
important role, so that we not only neglect loop corrections, but also set
$\lambda_3 = 0$ in this section.

\begin{figure}
	\centering
	\includegraphics[width=\textwidth]{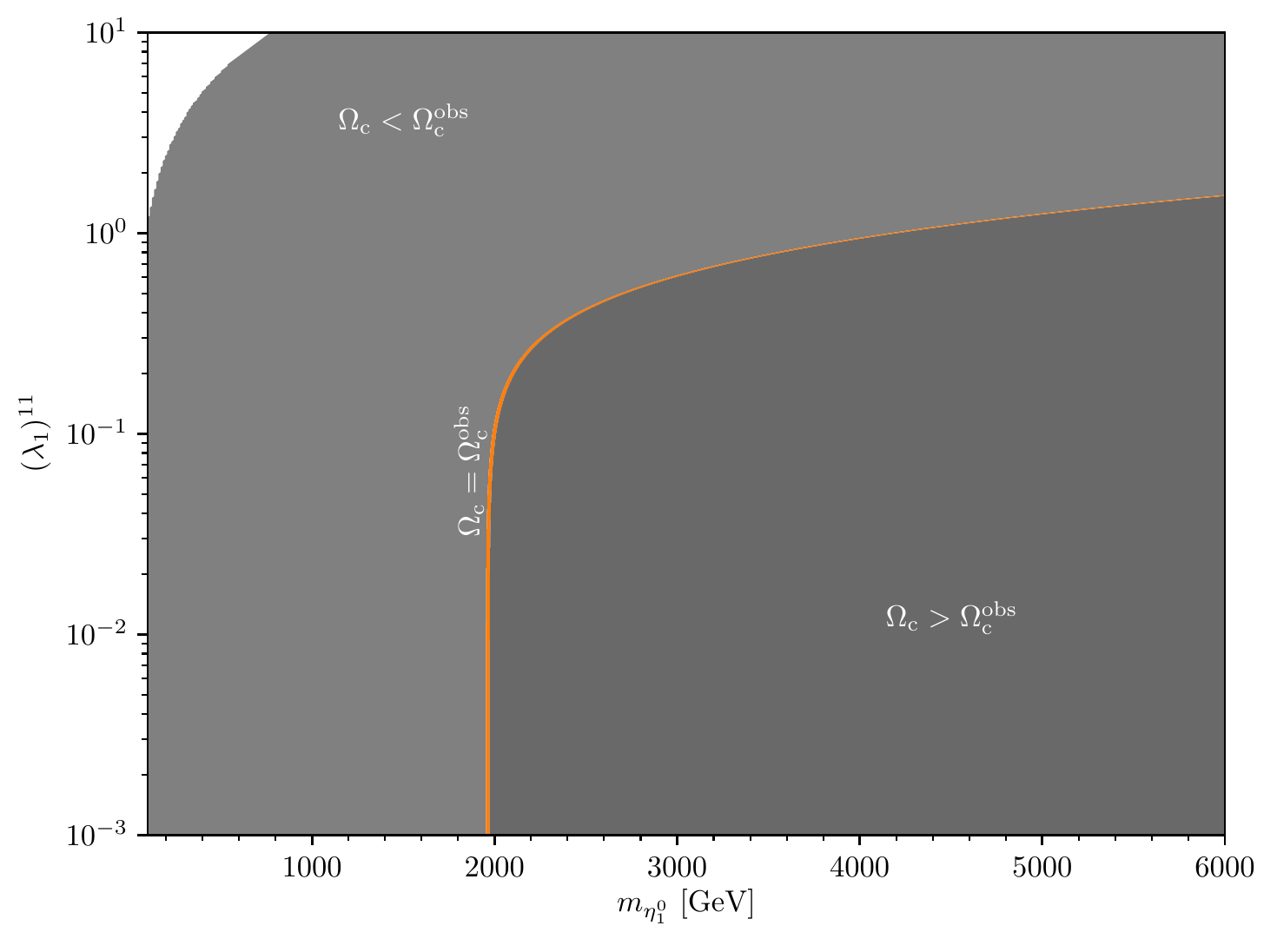}
	\caption{%
		The observed relic density contour ($\OmegacObs h^2$),
		separating regions of too small (above) and too large (below) relic density,
		as a function of the triplet scalar DM mass $m_{\eta_1^0}$ and its Higgs
		coupling $(\lambda_1)^{11}$. Here, we assume only one generation of scalars.%
	}
	\label{fig:triplet_scalar_relic}
\end{figure}

For one generation, the so-called inert triplet model was first studied in
ref.~\cite{Cirelli:2005uq} (although \emph{only} including gauge interactions)
with the result that DM had to be in a very narrow mass
region of \SI{2.0 +- 0.05}{\TeV} (see also ref.\ \cite{Hambye:2009pw,%
Fischer:2011zz,JosseMichaux:2012wj,Khan:2016sxm})
and that the spin-independent cross section was \SI{1.3e-9}{\pico\barn}.
For large couplings $\lambda_1 \simeq 1$ of DM to Higgs bosons,
a DM mass increase by about \SI{20}{\percent} was predicted due the opening of Higgs channels
in addition to the dominant DM annihilation into weak gauge bosons.
\Cref{fig:triplet_scalar_relic} confirms that
$m_{\eta_1^0} \simeq \SI{2}{\TeV}$ for $\lambda_1 \leq \num{0.2}$,
but also shows that the Higgs channels then become so efficient that the DM
mass rapidly increases into the multi-\si{\TeV} region (cf.\ figure~5 of
ref.~\cite{Hambye:2009pw} and figure~4 of ref.~\cite{JosseMichaux:2012wj}).
Qualitatively, this behaviour was
also found in figure~1 of ref.~\cite{Araki:2011hm}, but with a significant shift
to a larger DM mass of \SI{5.5}{\TeV}. Correcting the normalisation of the squared
neutral mass in their equation~(7) by a factor of two can partially explain this
discrepancy. With the same (incorrect) normalisation, an even larger DM mass of
\SI{7}{\TeV} was found, but not commented upon in ref.~\cite{Ayazi:2015mva}.
Surprisingly, and despite using more recent nuclear form factors, we can reproduce
the spin-independent detection cross sections in figure~2 of ref.~\cite{Araki:2011hm},
which also agree roughly with the numbers given in ref.\
\cite{Cirelli:2005uq,Hambye:2009pw} for sufficiently small $\lambda_1 \simeq
\num{0.01}$. As described there, gauge interactions increase the cross section at
one loop by \SI{0.86e-9}{\pico\barn}, while the Sommerfeld effect induces a
shift towards slightly larger DM masses (cf.\ figure~6 of ref.~\cite{Hambye:2009pw}).

\section{Radiative neutrino masses}
\label{sec:radiative_m_nu}

\begin{figure}
	\centering
	\includegraphics{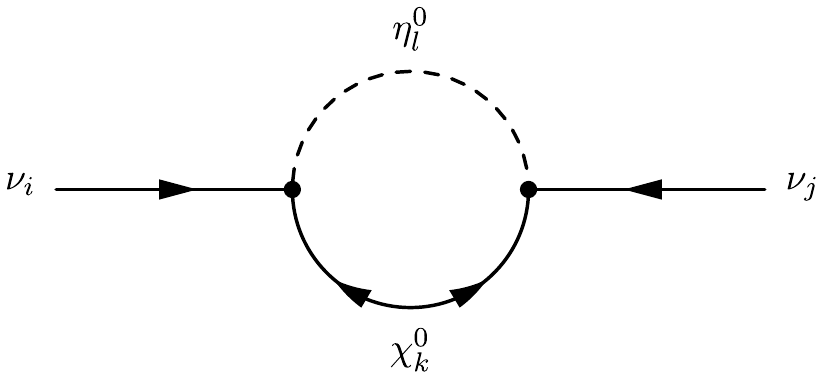}
	\caption{%
		Feynman diagram for the radiative generation of neutrino masses at
		the one-loop level after electroweak symmetry breaking and mixing ($k \in \{1, \dots, n_{\text{f}}\}$, $l \in \{1, \dots, n_{\text{s}}\}$).
		The convention for Feynman diagrams is the same as in ref.~\cite{Dreiner:2008tw}.%
	}
	\label{fig:feynman_diagram_m_nu}
\end{figure}

After electroweak symmetry breaking, neutrino masses in our model arise from
the one-loop diagram shown in \cref{fig:feynman_diagram_m_nu}. Only the $n_{\text{s}}$
neutral scalar and $n_{\text{f}}$ neutral fermion fields contribute to mass
generation, whereas the charged fields enter only into the propagator
correction. In our model, the $n_{\text{f}} = 3$ fermions are superpositions of $\SUIIL$
singlets and doublets, while the $n_{\text{s}} = 2$ scalars are superpositions of the
two generations of scalar triplets required for two non-zero neutrino masses.
Evaluating the two-point function in dimensional regularisation and summing
over all contributions leads to
\begin{align}
 (M_\nu)_{ij}
 &= \frac{1}{32\pi^2} \sum_{l=1}^{n_{\text{s}}} \lambda_6^{im} \lambda_6^{jn} (O_\eta)_{lm} (O_\eta)_{ln}
 \sum_{k=1}^{n_{\text{f}}} {(U_\chi)^*_{k3}}^2 \frac{m_{\chi_k^0}^3}{m_{\eta^0_l}^2 - m_{\chi_k^0}^2} \ln\lr \frac{m_{\chi_k^0}^2}{m_{\eta^0_l}^2} \rr
 \nonumber \\
 &= \frac{1}{32\pi^2} \sum_{l=1}^{n_{\text{s}}} A_l \lambda_6^{im} \lambda_6^{jn} (O_\eta)_{lm}(O_\eta)_{ln},
 \label{eq:neutrino_masses}
\end{align}
where the ultraviolet divergences have cancelled as expected for a
renormalisable model with vanishing neutrino masses at tree level.
As evident from \cref{eq:neutrino_masses}, the structure of the mass matrix is
chiefly determined by the number of scalar generations, which must be at least
as large as the desired number of non-zero neutrino masses.

\begin{figure}
	\centering
	\includegraphics[width=\textwidth]{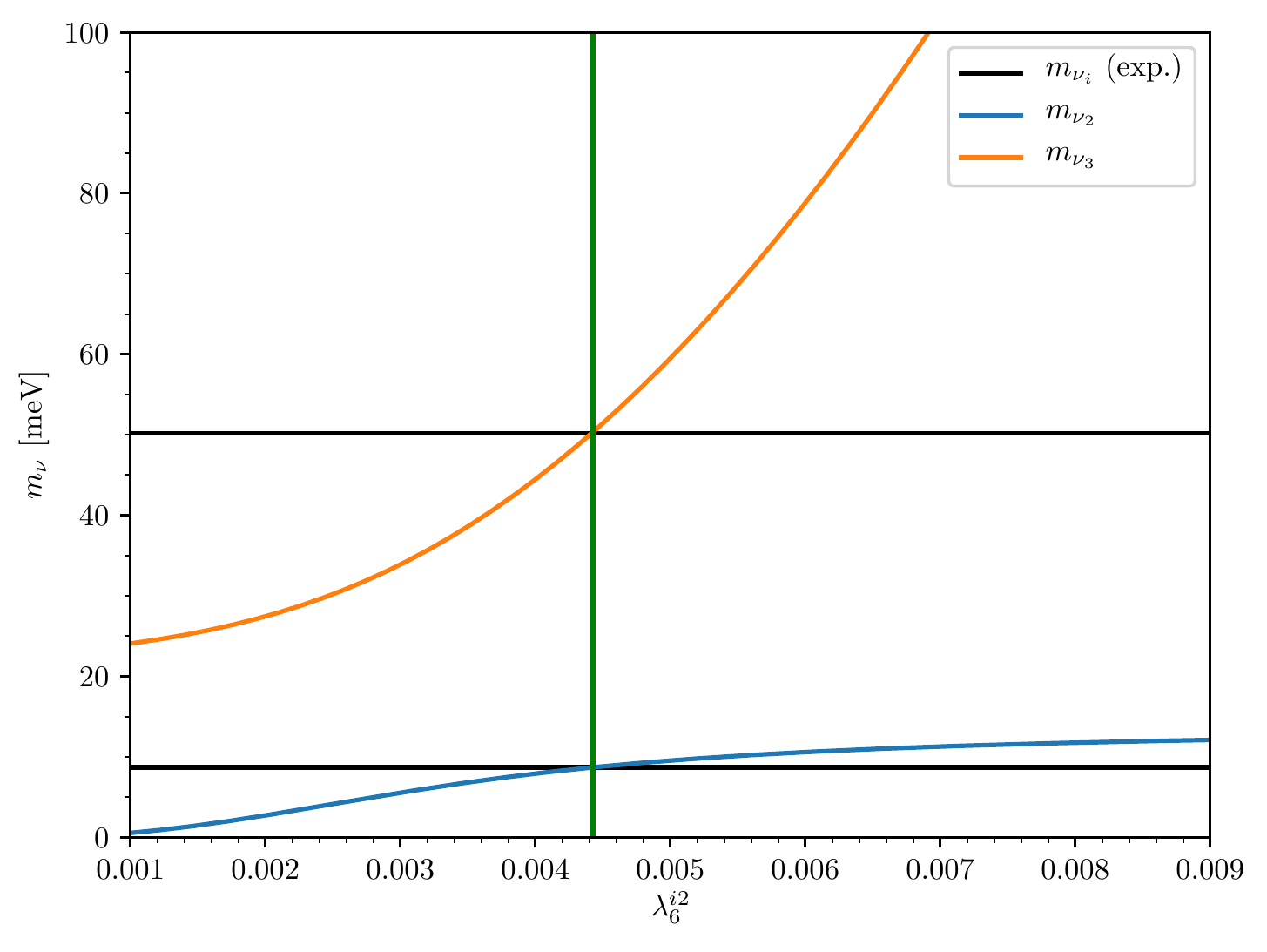}
	\captionof{figure}{%
		Neutrino masses $m_{\nu_i}$ in the model T1-3-B ($\alpha = 0$)
		with two generations of scalar triplets as a function of $\lambda_6^{i2}$.
		The used parameters are given in \cref{tab:neutrino_masses_2g_l6_input}.
		The experimentally determined neutrino masses are reproduced for
		$\lambda_6^{i2} = \num{0.0044}$.%
	}
	\label{fig:neutrino_masses_2g_l6}

	\bigskip

	\centering
	\renewcommand*{\arraystretch}{1.1}
	\captionof{table}{%
		Mass parameters and couplings used in \cref{fig:neutrino_masses_2g_l6}.%
	}
	\label{tab:neutrino_masses_2g_l6_input}
	\medskip
	\begin{tabular}{l r}
		\toprule
		Parameter & Value
		\\
		\midrule
		$(M_\phi)^{11}$ & \SI{1.4}{\TeV}
		\\
		$(M_\phi)^{22}$ & \SI{3}{\TeV}
		\\
		$(M_\phi)^{12}$ & $0$
		\\
		$M_\Psi$ & \SI{1}{\TeV}
		\\
		$M_{\psi\psi'}$ & \SI{1.5}{\TeV}
		\\
		$\lambda_1$, $\lambda_3$ & $0$
		\\
		$\lambda_4$, $\lambda_5$ & \num{5e-3}
		\\
		$\lambda_6^{i1}$ ($i \ne 3$) & \num{3e-4}
		\\
		$\lambda_6^{3l}$ & \num{3.05e-3}
		\\
		\bottomrule
	\end{tabular}
\end{figure}

The symmetric neutrino mass matrix $M_\nu$ in \cref{eq:neutrino_masses}
is diagonalised by the Pontecorvo-Maki-Nakagawa-Sakata (PMNS) matrix $U_\nu$,
\begin{equation}
	U_\nu^\dagger M_\nu U_\nu
	= D_\nu
	= \diag(\pm m_{\nu_1}, \pm m_{\nu_2}, \pm m_{\nu_3}).
	\label{eq:pmns}
\end{equation}
Since the neutrino mass matrix is of the same form as in model T1-3-A with
hypercharge parameter $\alpha = 0$ \cite{Klasen:2016vgl}, it can similarly be
expanded in the limit of small Yukawa couplings $\lambda_4, \lambda_5 \ll 1$.
At leading order, we obtain
\begin{equation}
 M_\nu \approx \SI{100}{\meV}\ \frac{M_\Psi}{\SI{1}{\TeV}}\
 \lr \frac{\lambda_6^{ij} \lambda_{4,5}}{\num{e-5}} \rr^2,
\end{equation}
i.\,e.\ for \si{\meV}-scale neutrinos only the product $\lambda_6^{ij} \lambda_{4,5}$
has to be of order \num{e-5}.
As an example, we show in \cref{fig:neutrino_masses_2g_l6} the
dependence of the two non-zero neutrino masses $m_{\nu_i}$ ($i \in \{2, 3\}$) on
the scalar-fermion coupling $\lambda_6^{i2}$.
The mass parameters and other couplings used in this figure are listed in
\cref{tab:neutrino_masses_2g_l6_input}.

The neutrino mass matrix $M_\nu$ depends explicitly on the couplings
\begin{equation}
	\lambda_6
	= \begin{pmatrix}
		\lambda_6^{e 1} & \lambda_6^{e2} \\
		\lambda_6^{\mu 1} & \lambda_6^{\mu 2} \\
		\lambda_6^{\tau 1} &\lambda_6^{\tau 2}
	\end{pmatrix}
\end{equation}
and on the masses $m_{\eta^0_l}$ and $m_{\chi^0_k}$ of the neutral scalars
and fermions, while the dependence on the other couplings $\lambda_{1,4,5}$
remains implicit in the mixing matrices. The Casas--Ibarra parametrisation
\cite{Casas:2001sr} then allows to obtain $\lambda_6$ from the experimental
neutrino data, once the other couplings and masses have been fixed. The
procedure is as follows: First, one factorises the explicit dependence on the
couplings $\lambda_6$ from the neutrino mass matrix in \cref{eq:neutrino_masses},
\begin{equation}
	M_\nu
	= \lambda_6 M \lambda_6^\tran
	= \lambda_6 U_M^\dagger D_M U_M \lambda_6^\tran,
\end{equation}
where in the last step the $n_{\text{s}} \times n_{\text{s}}$ matrix $M$ has been diagonalised to the matrix
$D_M$ using the unitary (here: orthogonal, as we have neglected complex
phases) matrix $U_M$. On the other hand, the neutrino mass matrix can be
diagonalised using the PMNS matrix as in \cref{eq:pmns}. Combining
these expressions, one obtains
\begin{equation}
	D_\nu
	= U_\nu^\dagger \lambda_6 U_M^\dagger D_M U_M \lambda_6^\tran U_\nu
\end{equation}
or, equivalently ($m_{\nu_1} = 0$),\footnote{%
	For a diagonal matrix $D$,
	$D^{\pm \frac{1}{2}} = \diag\Big(D_{11}^{\pm \frac{1}{2}}, D_{22}^{\pm \frac{1}{2}}, \dots\Big)$.%
}
\begin{equation}
	\diag(0, 1, 1)
	= D_\nu^{-\frac{1}{2}} U_\nu^\dagger \lambda_6 U_M^\dagger D_M U_M \lambda_6^\tran U_\nu D_\nu^{-\frac{1}{2}}.
\end{equation}
The left-hand side can be rewritten as $\diag(0, 1, 1) = R^\dagger R$ with
the \enquote{rotation matrix}
\begin{equation}
	R = \begin{pmatrix}
		0 & \cos(\theta) & \sin(\theta) \\
		0 & -\sin(\theta) & \cos(\theta)
	\end{pmatrix},
\end{equation}
which is thus given by
\begin{equation}
	R
	= D_M^{\frac{1}{2}} U_M \lambda_6^\tran U_\nu D_\nu^{-\frac{1}{2}}.
\end{equation}
Inverting this equation then leads to the Casas--Ibarra parametrisation
\begin{equation}
	\lambda_6 = U_\nu^* D_\nu^{\frac{1}{2}} R^\tran D_M^{-\frac{1}{2}} U_M^*.
	\label{eq:casas_ibarra}
\end{equation}
The couplings $\lambda_6$ of the dark scalar and fermion sectors are therefore
partially determined by the other couplings and masses, encoded in $U_M$ and
$D_M$, and by the measured neutrino mass differences (here: masses, since we
assume $m_{\nu_1} = 0$ and normal hierarchy) $D_\nu$ and mixing angles in $U_\nu$
\cite{Gonzalez-Garcia:2014bfa}, but they can still vary with the arbitrary
angle $\theta$ in the rotation matrix $R$.


\section{Dark matter direct detection and lepton flavour violation}
\label{sec:dd_lfv}

In this section, we now connect singlet--doublet fermions with triplet scalars
using the couplings $\lambda_6$ with the aim to not only explain the
observed small neutrino masses as described in the previous section, but also
in order to study the effect of the neutrino mass constraints on the nature,
allowed parameter space and direct detection prospects of the two DM candidates
in this combined model. In addition, we make predictions for the automatically
generated lepton flavour violating (LFV) processes and investigate the
corresponding current and future experiments with respect to their potential
to further restrict the viable parameter space of the model.

\begin{figure}
	\centering
	\includegraphics[width=\textwidth]{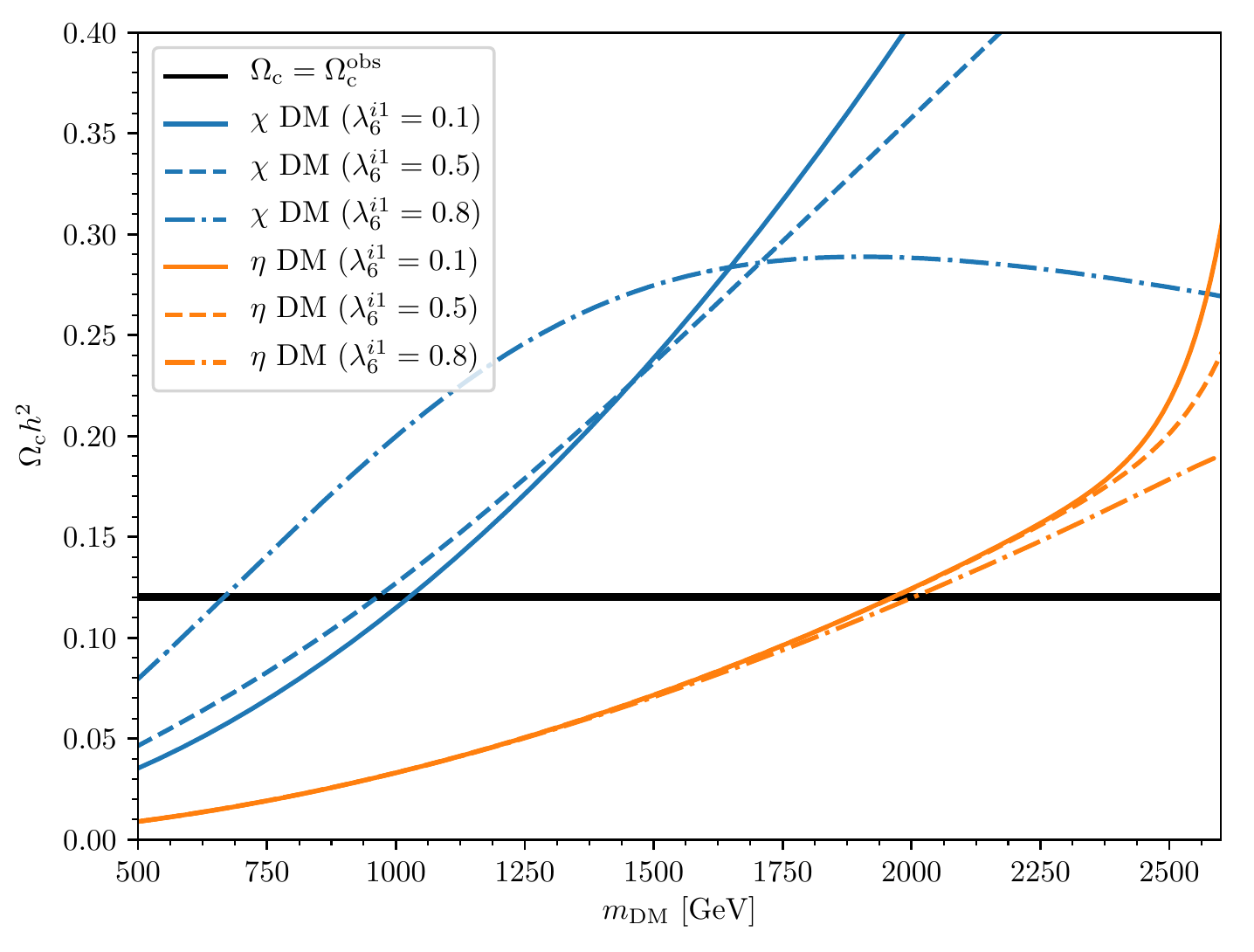}
	\captionof{figure}{%
		The relic density as a function of the DM mass for singlet--doublet
		fermion ($\chi^0_1$) and triplet scalar ($\eta^0_1$) DM.
		The used parameters are given in \cref{tab:t13b0_relic_mass_input}.%
	}
	\label{fig:t13b0_relic_mass}

	\bigskip

	\centering
	\renewcommand*{\arraystretch}{1.1}
	\captionof{table}{%
		Mass parameters and couplings used in \cref{fig:t13b0_relic_mass}.%
	}
	\label{tab:t13b0_relic_mass_input}
	\medskip
	\begin{tabular}{l r r}
		\toprule
		Parameter & Fermion DM & Scalar DM
		\\
		\midrule
		$(M_\phi)^{11}$ & \SI{3}{\TeV} & $[\SI{0.5}{\TeV}, \SI{3.5}{\TeV}]$
		\\
		$(M_\phi)^{22}$ & \SI{1000}{\TeV} & \SI{1000}{\TeV}
		\\
		$(M_\phi)^{12}$ & $0$ & $0$
		\\
		$M_\Psi$ & $[\SI{0.5}{\TeV}, \SI{2.8}{\TeV}]$ & \SI{3}{\TeV}
		\\
		$M_{\psi\psi'}$ & $M_\Psi$ & \SI{2.8}{\TeV}
		\\
		$\lambda_1$, $\lambda_3$, $\lambda_4$, $\lambda_5$
		& \num{e-2} & \num{e-2}
		\\
		$\lambda_6^{i2}$ & $0$ & $0$
		\\
		\bottomrule
	\end{tabular}
\end{figure}

For simplicity, we start in \cref{fig:t13b0_relic_mass} with the
dependence of the relic density on $\lambda_6^{i1}$, i.\,e.\ for only one generation
of scalars, assuming the couplings to be equal for all three lepton families.
The
results in \cref{fig:t13b0_relic_mass} therefore do not yet yield the
correct neutrino masses, but are shown for illustrative purposes only. The
mass parameters and other couplings used in this figure are listed in
\cref{tab:t13b0_relic_mass_input}.
As described in \cref{sec:dm_relic}, two distinct DM candidates give the
observed relic density: singlet--doublet fermions $\chi_1^0$ with masses up
to about \SI{1}{\TeV}, and triplet scalars with masses around \SI{2}{\TeV}. As we can now
see, this observation holds also in the case that the masses in the other
sector are only larger by \SIrange{0.2}{2.5}{\TeV}. Only for small mass splittings and
very large couplings $\lambda_6^{i1}$, fermion-scalar coannihilation processes can
reduce the relic density, as one can clearly see for triplet scalar DM,
which is otherwise insensitive to variations of $\lambda_6^{i1}$. In the heavy
scalar mass region, annihilation into weak bosons becomes ineffective, while
the one into Higgs bosons remains small due to $\lambda_1 = \num{0.01}$, so that
coannihilations can have a significant effect. For fermion DM, it is even more
pronounced and starts already at a mass of \SI{1.65}{\TeV}, whereas below this
value the effect is the opposite and small except for very large
$\lambda_6^{i1} = \num{0.8}$.

In the following two subsections, we present our main results of a numerical
random scan in the parameter space that
remains when the observed neutrino mass and mixing
constraints as well as the Higgs boson mass ($\pm \SI{2.5}{\GeV}$) are imposed.
In particular, we vary the couplings $\lambda_{1,4,5}$
in the range $[10^{-6}, 1]$ and the parameters $M_\phi$, $M_\Psi$ and
$M_{\psi\psi'}$ in the range $[\SI{10}{\GeV}, \SI{10000}{\GeV}]$.\footnote{%
	\Crefrange{fig:fermion_dm_direct_det_lfv}{fig:lfv_correlation_scalar}
	show the combined results of two random scans:
	an exploratory one where the mass parameters $M_\phi$, $M_\Psi$,
	$M_{\psi\psi'}$ were varied in the range $[\SI{10}{\GeV}, \SI{2000}{\GeV}]$
	and one using the full range given above.%
}
We also add a random sign for
the couplings $\lambda_{1,5}$ while keeping $\lambda_{4}$ always positive
\cite{Cheung:2013dua}.
All entries of $\lambda_1$ were set to the same value for simplicity (the off-diagonal element only introduces additional mass mixing to the scalar triplets).
The coupling $\lambda_3$ has no phenomenological implications, and the
couplings $\lambda_6$ are obtained from the Casas--Ibarra parametrisation
as a function of the other parameters. The free parameter $\theta$ to determine
$\lambda_6$ is taken randomly in the range $[0, 2\pi]$. Except for the
Casas--Ibarra angle $\theta$, the values of all parameters have been generated
from the exponential of randomly-generated numbers with a uniform distribution in the appropriate ranges.

Beyond the results presented in the following, we also checked the impact of electroweak precision data in the form of the $T$ parameter \cite{Tanabashi:2018oca} on the model's parameter space.
$T$ is defined as exactly zero in the SM and is given there (at tree level) by the expression
\begin{equation}
	\alpha T
	= \rho - 1
	= \frac{m_W^2}{m_Z^2 \cos(\theta_{\text{w}})}.
\end{equation}
For the points which survived all selection criteria (\cref{fig:fermion_dm_direct_det_lfv,fig:scalar_dm_direct_det_lfv}), \texttt{SPheno} was used to calculate $T$ to one-loop level.
The result is that the vast majority of points yields $T < \num{e-20}$, with a maximum value of $T = \num{1.2e-3}$.
Compared to the current experimental value of $T = \num{0.07 +- 0.12}$ \cite{Tanabashi:2018oca}, our results are orders of magnitude below the current bound and therefore $T$ does not currently restrict the parameter space.

\subsection{Singlet--doublet fermion dark matter}

\begin{figure}
	\centering
	\includegraphics[width=\textwidth]{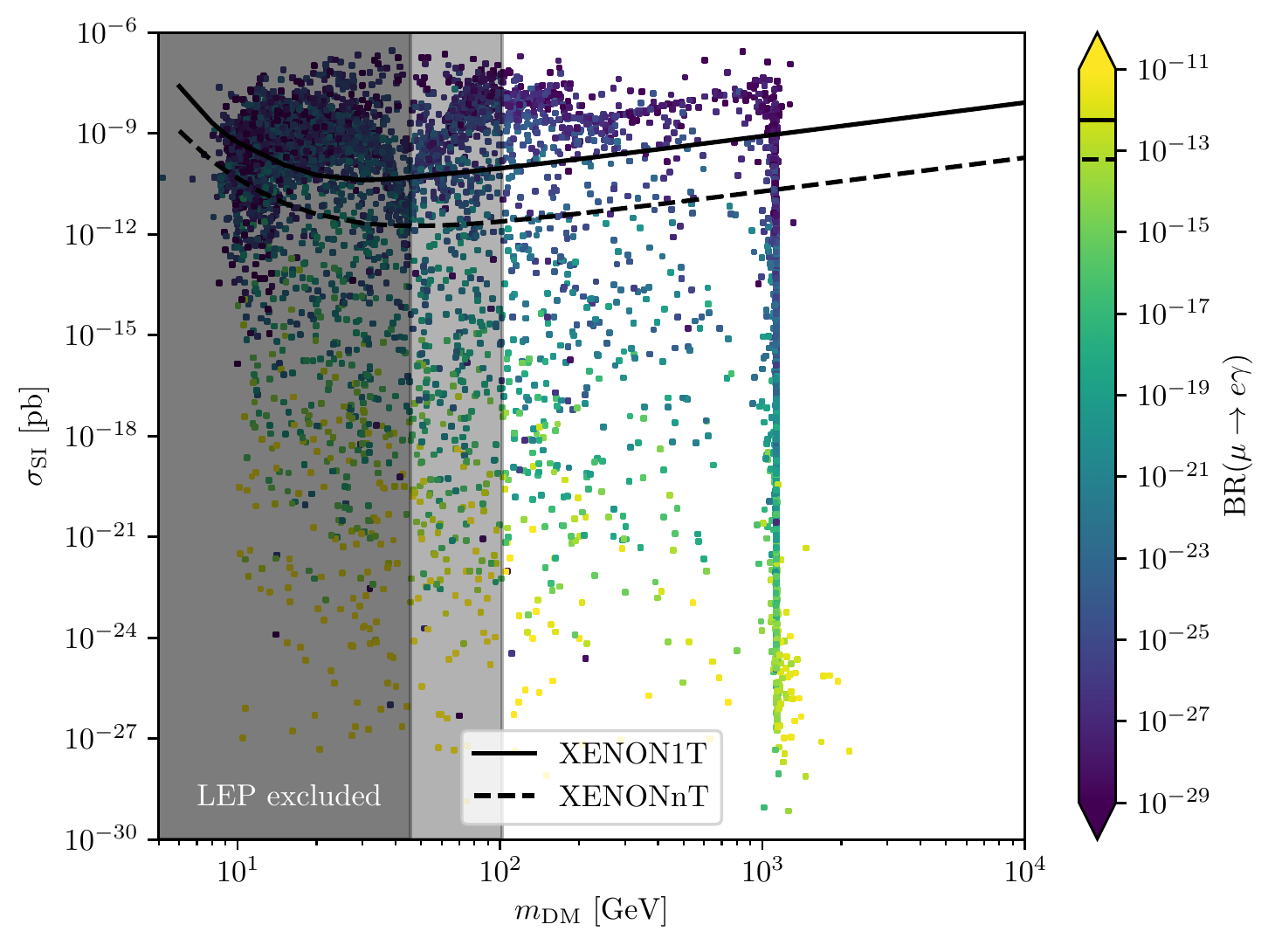}
	\caption{%
		The spin-independent direct detection cross section as a function
		of the DM mass for singlet--doublet fermion DM. The colours show the branching
		ratios for the LFV process $\mu \to e \gamma$.
		Also shown are the LEP limits on light neutral and charged particles (shaded areas) as well as current (full lines) and future (dashed lines) exclusion limits for the DM relic density from XENON1T \cite{Aprile:2018dbl} and XENONnT \cite{Aprile:2015uzo,oberlack}, and for $\mu \to e \gamma$ \cite{Adam:2013mnn,Baldini:2013ke}.%
	}
	\label{fig:fermion_dm_direct_det_lfv}
\end{figure}

About one third of all models with the observed neutrino masses and mixings feature singlet--doublet fermion DM, but only
a fraction of order \SI{0.02}{\percent} yield the correct DM relic density
$\OmegacObs h^2 = \num{0.120 +- 0.001}$ and Higgs mass.
These models are shown in \cref{fig:fermion_dm_direct_det_lfv} as a function of the DM
mass, together with their spin-independent direct detection cross
section and the branching ratio for the usually most sensitive LFV process
$\mu \to e \gamma$. The models accumulating at a DM mass of about \SI{1}{\TeV}
feature mostly doublet fermions, whereas lighter fermionic DM is generally
a superposition of singlet and doublet (cf.\ \cref{sec:dm_relic}).
A large doublet component below $\sfrac{m_Z}{2}$ (dark shaded area) is
excluded by the fact that the LEP measurement of the invisible $Z$ boson decay
width is consistent with three generations of active neutrinos
\cite{Tanabashi:2018oca}.
We assume this conservative exclusion limit independently of the size of the singlet--doublet mixing.
Furthermore, the accompanying, only slightly
heavier charged fermions are excluded below \SI{102}{\GeV} by largely
model-independent searches with the OPAL detector at LEP (light shaded area)
\cite{Abbiendi:2003yd}. The LHC limits for heavy long-lived charged particles
from ATLAS and CMS reach currently up to \SIlist{440; 490}{\GeV}, respectively, but
are more model-dependent
\cite{ATLAS:2014fka,Khachatryan:2016sfv}.
LHC signatures and constraints, also from indirect DM detection, up to a few
hundred GeV on singlet--doublet DM without triplet fermions have been discussed in \cite{Calibbi:2015nha}.
The spin-independent direct
detection cross section is compared to the current XENON1T exclusion limit
(full line) \cite{Aprile:2018dbl} and the expectation for 20 ton-years with
the XENONnT experiment (dashed line) \cite{Aprile:2015uzo}, which was
extrapolated linearly above \SI{1}{\TeV} \cite{oberlack}. XENON1T excludes most of the
models with small scalar-fermion couplings $\lambda_6$ and therefore also
little LFV. These models are therefore similar to those in the pure
singlet--doublet fermion DM
model. The combination with the scalar sector opens up a considerable
parameter space of leptophilic DM, i.\,e.\ with nuclear recoil cross sections
way below even the expected XENONnT sensitivity. Interestingly, one observes
a strong complementarity with LFV experiments, which already probe the models
with the smallest spin-independent direct detection cross section
\cite{Adam:2013mnn}.

\begin{figure}
	\centering
	\includegraphics[width=\textwidth]{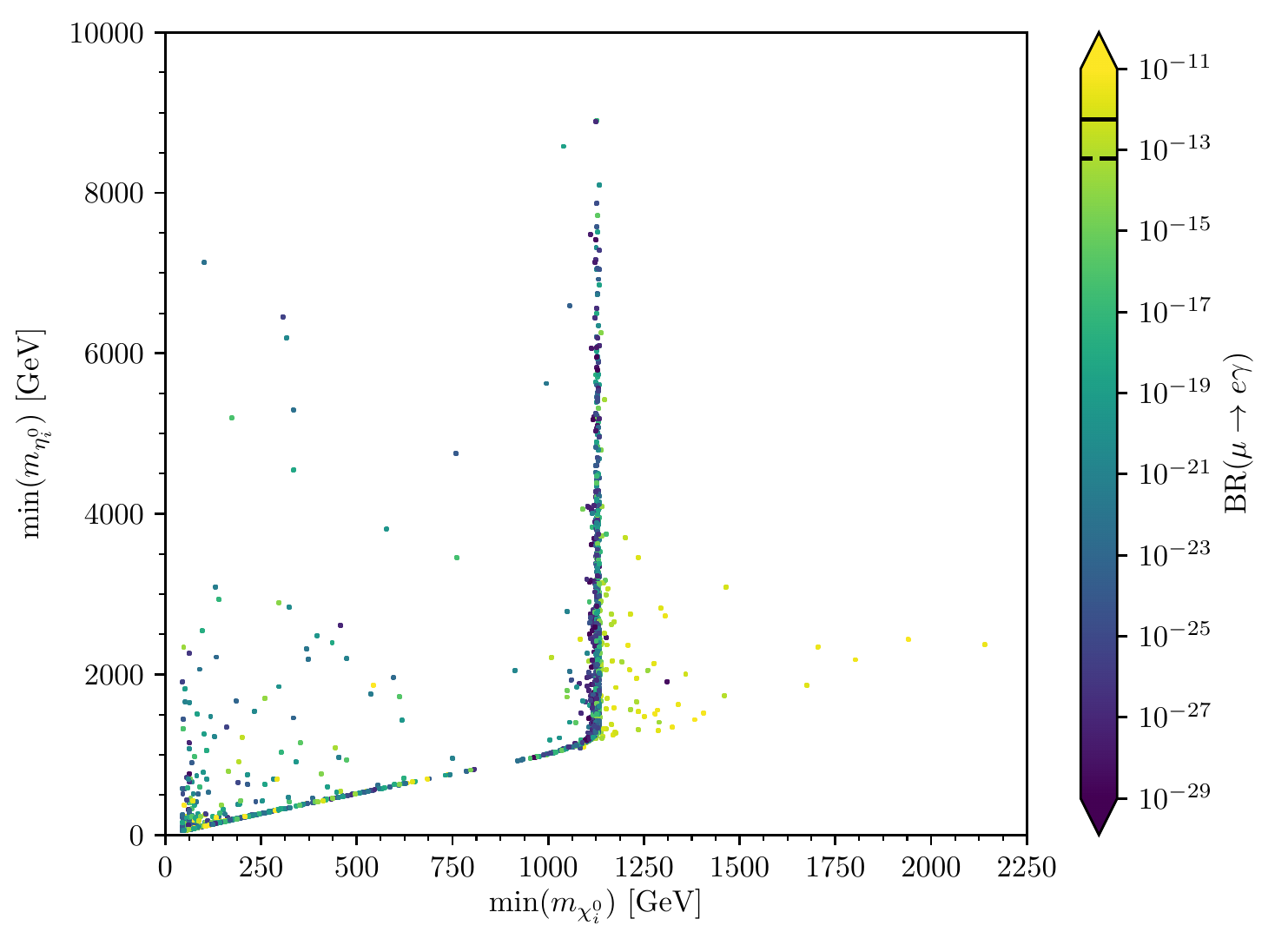}
	\caption{%
		The lightest neutral scalar vs.\ the lightest neutral fermion
		mass for viable models with singlet--doublet fermion DM. Also shown is the
		branching ratio for the LFV process $\mu \to e \gamma$ with current (full line) and future (dashed line) exclusion limits
		\cite{Adam:2013mnn,Baldini:2013ke}.%
	}
	\label{fig:fermion_dm_scalar_masses_lfv}
\end{figure}

\Cref{fig:fermion_dm_scalar_masses_lfv} shows the models that remain after imposing the XENON1T and
LEP limits of $\sfrac{m_Z}{2}$ for neutral scalars \cite{Tanabashi:2018oca} and
\SI{98}{\GeV} for charged scalars \cite{Abbiendi:2003yd}
in the mass plane of the lightest neutral fermion (the DM particle
$\chi_1^0$) and scalar ($\eta_1^0$). For most models, we observe a near
degeneracy of the masses up to \SI{1.1}{\TeV}, followed by a line of models with this
constant fermion mass and scalar masses up to the decoupling region. These
models will not be probed soon by the process $\mu \to e \gamma$
\cite{Baldini:2013ke}. Interestingly, there are a few models beyond a fermion
mass of \SI{1.1}{\TeV}, and these are in fact already excluded by LFV.

\begin{figure}
	\centering
	\includegraphics[width=\textwidth]{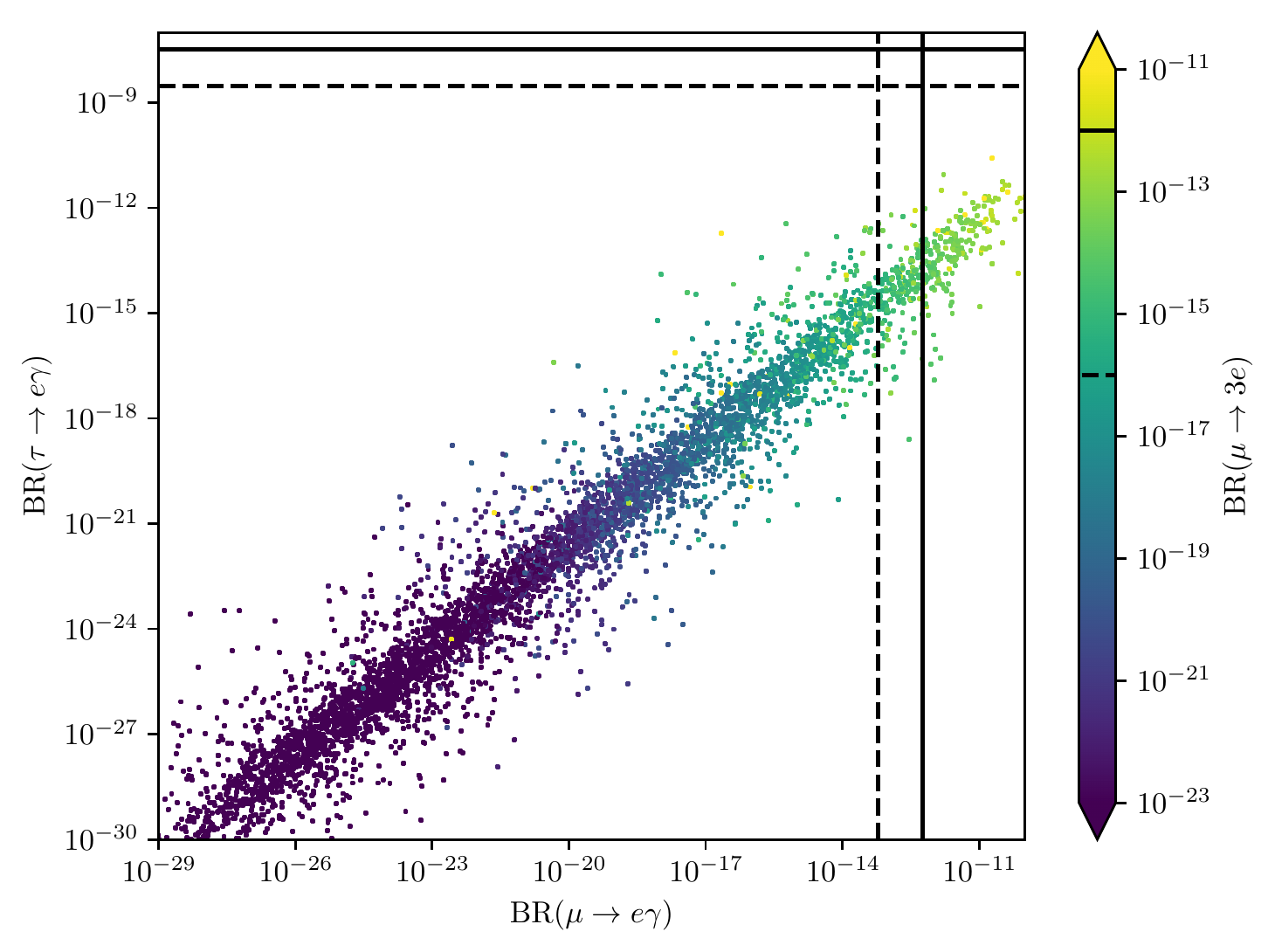}
	\caption{%
		Correlations of the branching ratios for the LFV processes
		$\mu \to e \gamma$, $\mu \to 3e$ and $\tau \to e \gamma$ for viable models
		with singlet--doublet fermion DM. Also shown are current (full lines)
		and future (dashed lines) exclusion limits
		\cite{Adam:2013mnn,Baldini:2013ke,Bellgardt:1987du,Blondel:2013ia,Aubert:2009ag,Aushev:2010bq}.%
	}
	\label{fig:lfv_correlation_fermion}
\end{figure}

In \cref{fig:lfv_correlation_fermion} we study which of the LFV processes $\mu \to e \gamma$,
$\mu \to 3e$ and $\tau \to e \gamma$ has the best prospects of further constraining
the remaining models. We observe that the three LFV processes are strongly
correlated, but that the process $\mu \to e \gamma$
\cite{Adam:2013mnn,Baldini:2013ke}, which is traditionally the most
sensitive process, is slightly outrivaled by its competitor $\mu\to3e$
\cite{Bellgardt:1987du,Blondel:2013ia}. In contrast,
the process $\tau\to e\gamma$ has currently no sensitivity to our model
\cite{Aubert:2009ag,Aushev:2010bq}.

\subsection{Triplet scalar dark matter}

\begin{figure}
	\centering
	\includegraphics[width=\textwidth]{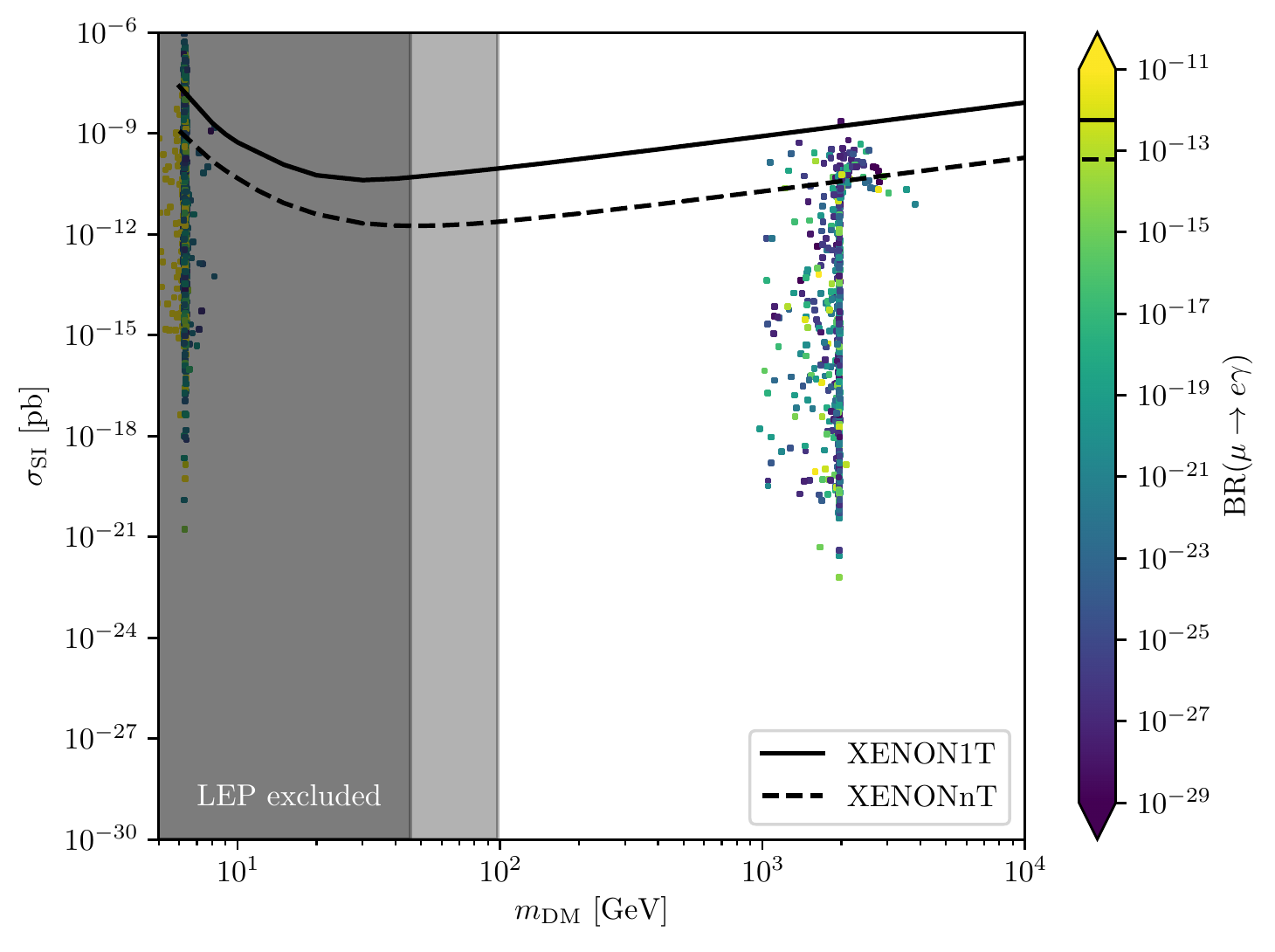}
	\caption{%
		The spin-independent direct detection cross section as a function
		of the DM mass for triplet scalar DM. The colours show the branching
		ratios for the LFV process $\mu \to e \gamma$.
		Also shown are the LEP limits on light neutral and charged particles (shaded areas) as well as current (full lines) and future (dashed lines) exclusion limits for the DM relic density from XENON1T \cite{Aprile:2018dbl} and XENONnT \cite{Aprile:2015uzo,oberlack}, and for $\mu \to e \gamma$ \cite{Adam:2013mnn,Baldini:2013ke}.%
	}
	\label{fig:scalar_dm_direct_det_lfv}
\end{figure}

About two thirds of all models with the observed neutrino masses and mixings feature triplet scalar DM, but only
a fraction of order \SI{0.02}{\percent} yield the correct DM relic density  $\OmegacObs h^2 = \num{0.120 +- 0.001}$ and Higgs mass.
These models are shown in \cref{fig:scalar_dm_direct_det_lfv} as a function of the DM
mass, together with their spin-independent direct detection cross
section and the branching ratio for the LFV process $\mu \to e \gamma$.
Similarly to the fermionic DM case, LEP constraints already rule out light DM candidates.
The precise measurements of the $Z$ width exclude charged scalars below $\sfrac{m_Z}{2}$ (dark shaded area) and, by virtue of the small mass splitting between charged and neutral components within the same triplet, we can assume this exclusion limit valid also for the neutral scalars, although they do not couple directly to the SM neutral boson because of their vanishing hypercharge.
Likewise, the exclusion limit from the aforementioned searches for charged scalars with the OPAL detector is adopted for the neutral components as well, ruling out light scalar DM below \SI{98}{\GeV} \cite{Abbiendi:2003yd} (light shaded area).
As for a pure triplet scalar model, we observe an accumulation of points
around a mass of \SI{2}{\TeV}. Many of these models have only very small couplings
$\lambda_6$ to the fermion sector and thus very little LFV. As $\lambda_1$
increases, so must the DM mass beyond \SI{2}{\TeV} to compensate for the stronger
Higgs annihilation. However, most of these models will soon be probed by
XENONnT, and those that will not can soon be excluded by the process
$\mu \to e \gamma$. While the mass region from \SIrange{1}{2}{\TeV} with leptophilic
fermion DM, that was opened up by coupling the fermion and scalar sectors, was
already excluded by LFV limits (see above), the corresponding models with
scalar DM are still allowed, but will soon be probed by the process
$\mu \to e \gamma$. Note that there exists in principle also a region of very
light triplet scalar DM of about \SI{6}{\GeV} mass, which is however excluded by
the LEP limits on light non-sterile neutral (dark shaded area) and charged
(light shaded area) particles.

\begin{figure}
	\centering
	\includegraphics[width=\textwidth]{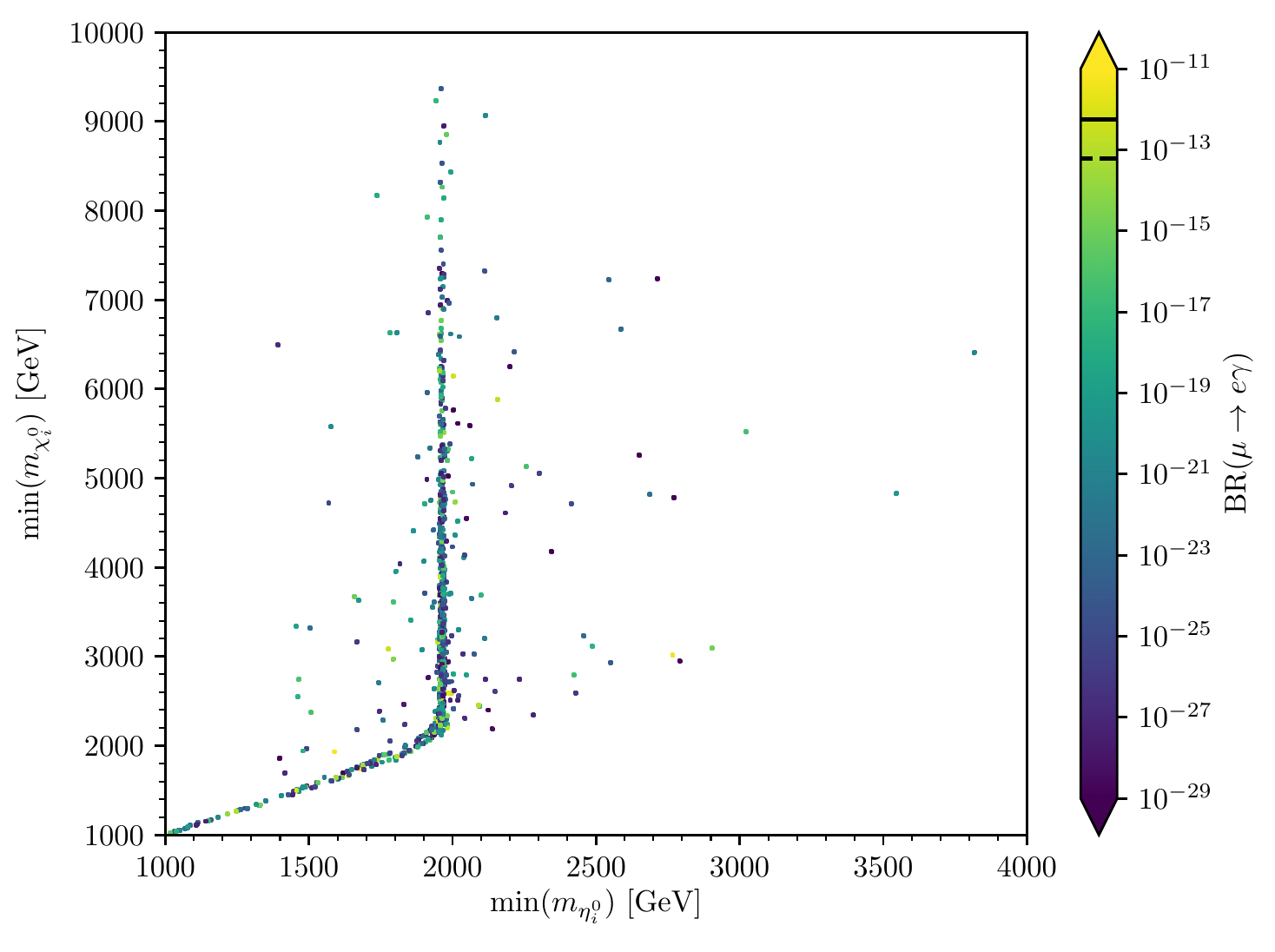}
	\caption{%
		The lightest neutral fermion vs.\ the lightest neutral scalar
		mass for viable models with triplet scalar DM.  Also shown is the
		branching ratio for the LFV process $\mu \to e \gamma$ with current (full line) and future (dashed line) exclusion limits
		\cite{Adam:2013mnn,Baldini:2013ke}
	}
	\label{fig:scalar_dm_fermion_masses_lfv}
\end{figure}

\Cref{fig:scalar_dm_fermion_masses_lfv} shows the models that remain after imposing the XENON1T and
LEP limits in the mass plane of the lightest neutral scalar (the DM particle
$\eta_1^0$) and fermion ($\chi_1^0$). As for fermion DM, we observe a near
degeneracy of the masses from \SIrange{1}{2}{\TeV}, i.\,e.\ the two competing DM candidates
remain close in mass, but change spin around \SI{1}{\TeV}. Despite the small mass
difference, coannihilation is rare due to the small couplings $\lambda_6$.
Once the triplet scalar mass has reached the favoured value of \SI{2}{\TeV}, the
fermion mass can increase up to the decoupling region. As discussed above,
models with higher scalar mass are possible, in particular when the Higgs
coupling $\lambda_1$ is large, but will soon be probed by either XENONnT
or LFV.

\begin{figure}
	\centering
	\includegraphics[width=\textwidth]{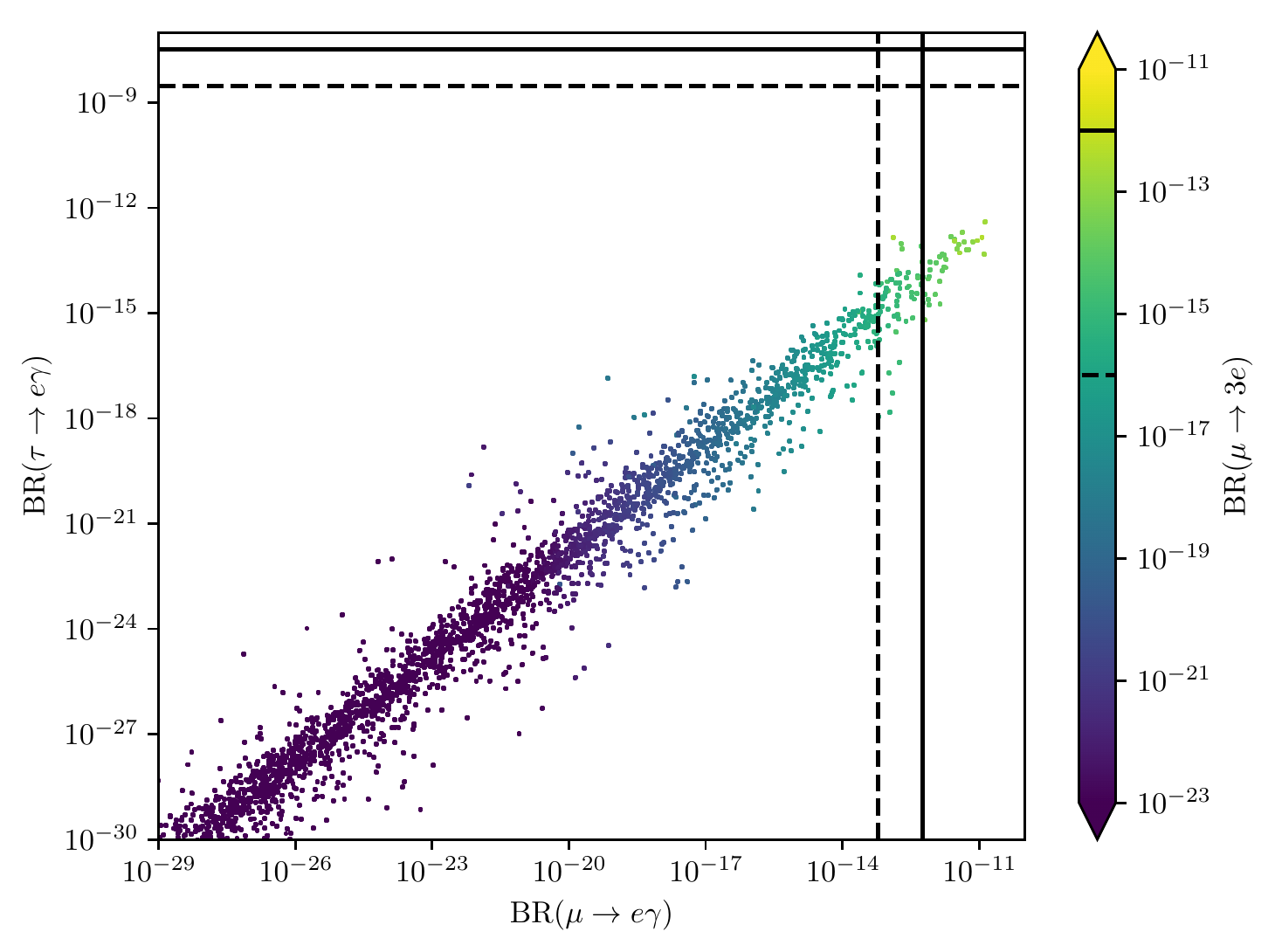}
	\caption{%
		Correlations of the branching ratios for the LFV processes
		$\mu \to e \gamma$, $\mu \to 3e$ and $\tau \to e \gamma$ for viable models
		with triplet scalar DM. Also shown are current (full lines)
		and future (dashed lines) exclusion limits
		\cite{Adam:2013mnn,Baldini:2013ke,Bellgardt:1987du,Blondel:2013ia,Aubert:2009ag,Aushev:2010bq}.%
	}
	\label{fig:lfv_correlation_scalar}
\end{figure}

Finally, we study the correlation of the three LFV processes $\mu \to e \gamma$,
$\mu \to 3e$ and $\tau \to e \gamma$ in \cref{fig:lfv_correlation_scalar} for the case of triplet
scalar DM. As for singlet--doublet fermion DM, we observe a strong correlation
of the three processes, a similar sensitivity of the two muon decay processes
with a small advantage for the one to three electrons, and no sensitivity for
the $\tau$ decay. Therefore, in contrast to the fermion DM case, the parameter
space for scalar DM can only be fully probed with additional experimental
effort.

\section{Conclusion}
\label{sec:conclusion}

In this paper, we have combined the singlet--doublet fermion model
with the triplet scalar model in order to explain not only the observed
DM relic density, but also the neutrino masses and mixings, which were
generated radiatively. This model allows in addition for the correct
Higgs boson mass, couplings of natural size, masses in the \si{\TeV} range
and gauge coupling unification at a scale of $\mathcal{O}(\SI{e13}{\GeV})$.

We first updated and clarified some discrepancies in the literature for
the two separate DM models, showing that the new XENON1T limits roughly
double the excluded parameter space for the singlet--doublet fermion model
and confirming that the triplet scalar DM mass is around \SI{2}{\TeV} for small
Higgs couplings, but reaches into the multi-\si{\TeV} region for large Higgs
couplings.

With two generations of scalars, we generated two non-zero neutrino masses
and implemented the experimental mass and mixing constraints using the
Casas--Ibarra parametrisation. Analytically, we found that the product of
Yukawa and new scalar-fermion couplings had to be of $\mathcal{O}(\num{e-5})$
for a DM mass of \SI{1}{\TeV}.

We found that DM in our model is fermionic up to the \si{\TeV} scale and scalar
beyond with small mass splittings not only among the partners of the weak
isospin multiplets, but also the lightest fermions and scalars. The
scalar-fermion couplings opened the parameter space, so that leptophilic
singlet--doublet fermion DM below \SI{1}{\TeV}, but above the LEP limits of $\sfrac{m_Z}{2}$
for active neutral fermions from the invisible $Z$ boson decay width and \SI{102}{\GeV}
from largely model-independent searches for charged fermions, became again
viable below the XENON1T
exclusion limit, as did triplet scalar DM between \SIlist{1; 2}{\TeV}. In both
regions, we observed an interesting complementarity between the expectations
for XENONnT and for LFV experiments. For the latter, the process $\mu \to 3e$
has the largest sensitivity, followed by $\mu \to e \gamma$, whereas or model
is insensitive to $\tau \to e \gamma$. For triplet scalar DM, we found a
viable region at masses of a few \si{\GeV}, which is, however, excluded by LEP
limits below $\sfrac{m_Z}{2}$ for neutral and \SI{98}{\GeV} for charged scalars. While
current LHC limits reach to considerably higher masses of \SIrange{440}{490}{\GeV},
they are generally more model-dependent. Triplet scalar DM models with
masses above \SI{2}{\TeV} require a large Higgs coupling leading to a large
spin-independent cross section that will soon be probed by XENONnT.

\section*{Acknowledgements}

This work has been supported by the BMBF under contract 05H18PMCC1 and the DFG
through the Research Training Group~2149 \enquote{Strong and weak interactions --
from hadrons to dark matter}. We thank U.\ Oberlack for communication on the
XENONnT expectations and C.\ Yaguna for helpful comments on the manuscript.

\bibliography{bib}

\end{document}